\documentclass{article}

\usepackage[preprint]{neurips_2023}
\usepackage[utf8]{inputenc} 
\usepackage[T1]{fontenc}    
\usepackage{hyperref}       
\usepackage{url}            
\usepackage{booktabs}       
\usepackage{amsfonts}       
\usepackage{nicefrac}       
\usepackage{microtype}      
\usepackage{xcolor}         
\usepackage{graphicx}
\usepackage{subfig} 
\usepackage{siunitx}
\usepackage{multirow}
\usepackage{makecell}

\usepackage[ruled]{algorithm2e}
\usepackage{algorithmicx}  
\usepackage{algpseudocode}  
\usepackage{amsmath}  

\renewcommand{\figurename}{Fig.} 

\title{scRNA-seq Data Clustering by Cluster-aware Iterative Contrastive Learning}

\author{%
  Weikang Jiang$^{1}$\thanks{Co-first authors who contribute equally to this work.} , Jinxian Wang$^{1}$\footnotemark[1], Jihong Guan$^{2}$, Shuigeng Zhou$^{1,}$\thanks{Corresponding author.} \\
  $^{1}$Shanghai Key Lab of Intelligent Information Processing, \\and School of Computer Science, 
Fudan University \\
  $^{2}$Department of Computer Science and Technology, Tongji University\\
  \texttt{\{sgzhou, 20210240030\}@fudan.edu.cn}  \\ 
  \texttt{wangjx22@m.fudan.edu.cn, jhguan@tongji.edu.cn } 
}

\begin{document}
\maketitle

\begin{abstract}
    Single-cell RNA sequencing~(scRNA-seq) enables researchers to analyze gene expression at single-cell level. One important task in scRNA-seq data analysis is unsupervised clustering, which helps identify distinct cell types, laying down the foundation for other downstream analysis tasks. 
    In this paper, we propose a novel method called \emph{Cluster-aware Iterative Contrastive Learning}~(CICL in short) for scRNA-seq data clustering, which utilizes an iterative representation learning and clustering framework to progressively learn the clustering structure of scRNA-seq data with a cluster-aware contrastive loss. CICL consists of a Transformer encoder, a clustering head, a projection head and a contrastive loss module. First, CICL extracts the feature vectors of the original and augmented data by the Transformer-encoder. Then, it computes the clustering centroids by K-means and employs the student’s t-distribution to assign pseudo-labels to all cells in the clustering head. The projection-head uses a Multi-Layer Perceptron~(MLP) to obtain projections of the augmented data.
    At last, both pseudo-labels and projections are used in the contrastive loss to guide the model training. Such a process goes iteratively so that the clustering result becomes better and better. Extensive experiments on 25 real-world scRNA-seq datasets show that CICL outperforms the state-of-the-art~(SOTA) methods. Concretely, CICL surpasses the existing methods by from 14\% to 280\%, and from 5\% to 133\% on average in terms of performance metrics ARI and NMI respectively.  
Source code is available at \href{https://github.com/Alunethy/CICL}{https://github.com/Alunethy/CICL}.
\end{abstract}

\section{Introduction}
Each cell possesses unique characteristics and biological functions defined by its gene transcription activities. Conventional bulk RNA sequencing measures the average transcription levels of a multitude of cells, thereby obscuring the heterogeneity among individual cells. In the past decade, the rapid progress of single-cell RNA sequencing~(scRNA-seq) technologies~\citep{tang2009mrna} enables transcriptome-wide gene expression measurement in individual cells, which greatly helps deepen our understanding of cellular heterogeneity and propels the research on cell biology, immunology, and complex diseases~\citep{van2019single}.
Identifying cell types is a fundamental step in unraveling complex biological processes such as cellular differentiation, lineage commitment, and gene regulation~\citep{deng2021single}. As such, cell clustering becomes an important task in scRNA-seq analysis. However, the inherent high-dimensionality, noise, and sparsity of scRNA-seq data present severe challenges for scRNA-seq data clustering analysis~\citep{kiselev2019challenges,stegle2015computational}.

Up to now, many models or algorithms have been developed for scRNA-seq data clustering. 
Early scRNA-seq clustering methods mainly rely on traditional dimensionality reduction and clustering methods. For example, pcaReduce~\citep{vzurauskiene2016pcareduce} combines PCA and K-means, iteratively merging cluster pairs based on related probability density function. Recognizing the importance of similarity metrics in the clustering task, SIMLR~\citep{wang2018simlr} amalgamates multiple kernels to learn sample similarity and perform spectral clustering. Seurat~\citep{satija2015spatial} employs a graph-based community detection algorithm, while Louvain~\citep{blondel2008fast} is based on the shared nearest neighbor graph to identify cell types.

In the past decade, with the rapid development of deep learning, deep neural networks~(DNN) have been extensively applied to scRNA-seq data clustering to address the limitations of conventional methods~\citep{flores2022deep}. DEC~\citep{xie2016unsupervised} and IDEC~\citep{guo2017improved}, based on autoencoders~(AE), use KL divergence as the clustering loss, achieving simultaneous learning of feature representations and cluster assignments. To address the pervasive dropout events in scRNA-seq data, DCA~\citep{eraslan2019single} proposes a zero-inflated negative binomial~(ZINB) model to better characterize the distribution of scRNA-seq data, and uses the negative likelihood as the reconstruction loss instead of the frequently-used mean-square error~(MSE) loss in autoencoders. scVI~\citep{lopez2018deep} is a deep generative model based on variational autoencoders, which can do various scRNA-seq data analyses such as data imputation, clustering, and visualization. scDeepCluster~\citep{tian2019clustering} introduces a novel model-based deep learning clustering approach. By combining the ZINB model with the DEC algorithm, it is designed to capture the underlying cluster structure of scRNA-seq data. scDHA~\citep{tran2021fast} exploits a stacked Bayesian self-learning network to learn compact and generalized representations of scRNA-seq data. 

\begin{figure*}[ht]
  \centering
 \includegraphics[width=0.9\textwidth]{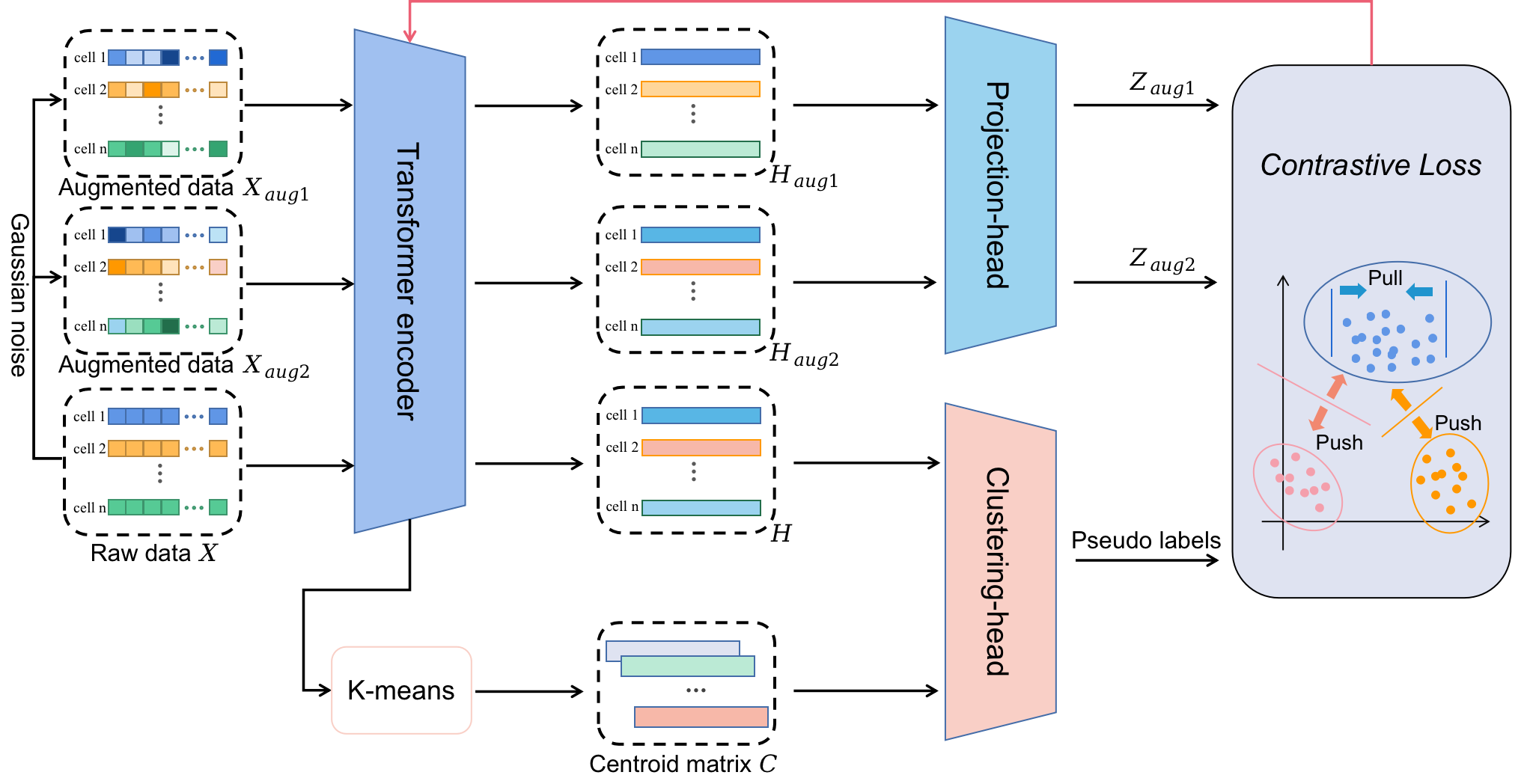}
 \vspace{-0.3cm}
  \caption[CICL]{
     The framework of CICL. We first generate two augmented views $X_{aug1}$ and $X_{aug2}$ of the raw data $X$ by adding Gaussian noise. Then, we obtain the representations $H$, $H_{aug1}$ and $H_{aug2}$ of $X$, $X_{aug1}$ and $X_{aug2}$ by a transformer encoder. 
     Next, we perform K-means on $H$ to get the centroid matrix $C$. After that, we feed $H$ and $C$ to the clustering-head and get a pseudo-label for each cell.
     Meanwhile, the projection-head encodes the $H_{aug1}$ and $H_{aug2}$ to obtain their projections $Z_{aug1}$ and $Z_{aug2}$.
     Finally, by using a novel contrastive loss, we align the positive pairs and contrast the negative pairs simultaneously. This process forms a loop to improve representations and clustering iteratively.}
  \label{fig:workflow}
  \vspace{-0.1cm}
\end{figure*}

To leverage the relationships between cells, some studies construct the cell-cell graph and apply Graph Neural Networks~(GNNs) to learn the representations of cells. scDSC~\citep{gan2022deep} formulates and aggregates cell-cell relationships with graph neural networks and learns latent gene expression patterns using a ZINB model based autoencoder. GraphSCC~\citep{zeng2020accurately} integrates the structural relationships between cells into scRNA-seq clustering by employing a graph convolutional network. It also utilizes a dual self-supervised module to cluster cells and guide the training process. Furthermore, 
Some other works have tried to train models using manual annotations as supervisory information or prior knowledge, as demonstrated in transfer learning and meta-learning methods~\citep{hu2020iterative}. While these methods can deliver excellent results on specific datasets, they also face serious challenge of scalability.

Contrastive learning~(CL) has been widely used in computer vision and natural language processing~\citep{chen2020simple,he2020momentum}. There have also been endeavors to incorporate contrastive learning into scRNA-seq data clustering. For instance, contrastive-sc~\citep{ciortan2021contrastive} proposes a contrastive learning based method for scRNA-seq data by masking a certain proportion of data features to obtain augmented data. Similar to most practices in contrastive learning, this method designates augmented pairs as positive samples, while considering all other pairs as negatives. scNAME~\citep{wan2022scname} improves the conventional contrastive loss by proposing a new neighborhood contrastive loss combined with an ancillary mask estimation task, characterizing feature correlation and pairwise cell similarity better. CLEAR~\citep{han2022self} employs multiple data augmentation methods to simulate different noise types, uses the infoNCE~\citep{oord2018representation} loss as a contrastive loss, and generates feature representations for scRNA-seq data with the momentum update strategy of encoder. However, these methods mainly apply the standard contrastive learning directly, failing to adapt the selection of positive and negative samples to the clustering task. 

This paper aims to boost the performance of scRNA-seq data clustering by exploring new methods. Our contributions are two-fold. On the one hand, we propose a \emph{Cluster-aware Iterative Contrastive Learning}~(CICL) method for scRNA-seq data clustering. CICL employs an iterative representation learning and clustering framework with a cluster-aware contrastive loss, it can progressively improve the clustering result by comprehensively exploiting the hidden cluster structure for scRNA-seq data representation. On the other hand, we conduct extensive experiments on 25 real-world datasets, which show that our method outperforms the SOTA methods in most cases.

\section{Materials and Methods}

\subsection{Datasets and Performance Metrics}\label{data_eva}
The proposed CICL method is evaluated on 25 real scRNA-seq datasets, and each dataset contains cells whose labels are known as prior or validated in the previous studies. The 25 datasets were derived from 7 different sequencing platforms. The smallest dataset contains only 90 cells, while the largest dataset has 48,266 cells. The number of cell subtypes in these datasets ranges from 2 to 15. Statistics of these datasets are presented in Table~\ref{dataset}.

\begin{table}[htbp]
\scriptsize
\vspace{-0.2cm}
\fontsize{9}{9}\selectfont
\caption{Datasets used in our experiments.}
\begin{center}
\begin{tabular}{|c|c|c|c|c|}
\hline
\textbf{Dataset}& \textbf{Platform} & \textbf{\#Cells} &\textbf{\#Genes} &\textbf{\#Subtypes}\\
\hline
Adam~\citep{adam2017psychrophilic} &Drop-seq &3660 &23797 &8 \\
\hline
Bach~\citep{bach2017differentiation} &10X &23184 &19965 &8 \\
\hline
Bladder~\citep{tabula2020single} & 10X &2432 &22966 &2 \\
\hline
deng~\citep{deng2014single} &Smart-seq2 &268 &22431 &6 \\
\hline
Diaphragm~\citep{tabula2020single} &10X &1858 &22966 &6 \\
\hline
hrvatin~\citep{hrvatin2018single} &Drop-seq&48266 &25187 &8 \\
\hline
Klein~\citep{klein2015droplet} & inDrop &2717 &24047 &4 \\
\hline
kolodziejczyk~\citep{kolodziejczyk2015single} & SMARTer &704 &38616 &3 \\
\hline
Limb\_Muscle~\citep{tabula2020single} & 10X&3855 &22966 &6 \\
\hline
Mammary\_Gland~\citep{tabula2020single} &10X&3282 &22966 &4 \\
\hline
muraro~\citep{muraro2016single} & CEL-seq2 &2126 &19127 &10 \\
\hline
Plasschaert~\citep{plasschaert2018single} & inDrop &6977 &28205 &8 \\
\hline
pollen~\citep{pollen2014low} & SMARTer &301 &23730 &11 \\
\hline
Qx\_Spleen~\citep{schaum2018single} &10X &9552 &23341 &5 \\
\hline
Qx\_Trachea~\citep{schaum2018single} &10X &11269 &23341 &5 \\
\hline
QS\_Diaphragm~\citep{schaum2018single} &Smart-seq2 &870 &23341 &5 \\
\hline
QS\_Heart~\citep{schaum2018single} &Smart-seq2 &4365 &23341 &8 \\
\hline
QS\_Limb\_Muscle~\citep{schaum2018single} & Smart-seq2&1090 &23341 &6 \\
\hline
QS\_Lung~\citep{schaum2018single} &Smart-seq2  &1676 &23341 &11 \\
\hline
QS\_Trachea~\citep{schaum2018single} &Smart-seq2 &1350 &23341 &4 \\
\hline
Romanov~\citep{romanov2017molecular} &SMARTer &2881 &21143 &7 \\
\hline
Tosches\_turtle~\citep{tosches2018evolution} & Drop-seq &18664 &23500 &15 \\
\hline
Wang\_lung~\citep{wang2018pulmonary} &10X&9519 &14561 &2 \\
\hline
yan~\citep{yan2013single} & Tang &90 &20214 &6 \\
\hline
Young~\citep{young2018single} &10X &5685 &33658 &11 \\
\hline
\end{tabular}
\end{center}
\label{dataset}
\vspace{-0.2cm}
\end{table}

We preprocess the 
scRNA-seq data with the Python package SCANPY~\citep{wolf2018scanpy}, following the strategy in~\citep{tian2019clustering}. Specifically, given the raw read counts (i.e., gene expression matrix), we first filter out cells and genes without counts. Then, we calculate the library size of each cell as the total number of read counts per cell, and obtain the size factor of each cell via dividing its library size by the median of all library sizes. Thirdly, we obtain the normalized read count by dividing the raw read count with the size factor of each cell, followed by a natural $log$ transformation. Furthermore, we consider only the top-$t$ highly variable genes according to their normalized dispersion values, and set $t$ to 500 by default in our paper. Finally, we transform the normalized read counts into z-score data. 

Two widely used metrics, normalized mutual information~(NMI) and adjusted rand index~(ARI) are used to evaluate clustering performance. 
NMI measures the similarity between the predicted labels and the real labels. Specifically, given the predicted labels $U=\left[u_{1}, u_{2}, ..., u_{N}\right] \in \mathbb{R}^{N} $ and the real labels $V=\left[v_{1}, v_{2}, ..., v_{N}\right] \in \mathbb{R}^{N} $, $N$ denotes the number of cells, NMI is evaluated as follows:
\begin{equation}
    NMI=\frac{I(U,V)}{max(H(U),H(V))}
\end{equation}
where $I(U,V)=\sum_{u}\sum_{v}p(u, v)log \frac{p(u, v)}{p(u)p(v)} $ calculates the mutual information between $U$ and $V$, $p(u, v)$ is the joint distribution of $U$ and $V$, $p(u)$ and $p(v)$ are marginal distributions respectively. $H(U) = \sum_{u} p(u)log(p(u))$ is the entropy of clustering $U$. Similarly, $H(V) = \sum_{v} p(v)log(p(v))$. 

ARI was also used to measure the similarity between clustering results and true categories. It solves the problem of insufficient punishment of RI and considers the impact of random assignments. The value of ARI ranges from -1 to 1. The larger the value, the more similar the clustering result is to the real categories. ARI is defined as
\begin{equation}
    ARI=\frac{\begin{matrix} \sum_{ij} {n_{ij} \choose 2} \end{matrix}-[
    \begin{matrix} \sum_{i} {a_i \choose 2} \end{matrix} \begin{matrix} \sum_{j} {b_j \choose 2} \end{matrix}
    ]/{N \choose 2}}{[
    \begin{matrix} \sum_{i} {a_i \choose 2} \end{matrix} \begin{matrix} \sum_{j} {b_j \choose 2} \end{matrix}
    ]/2-[
    \begin{matrix} \sum_{i} {a_i \choose 2} \end{matrix} \begin{matrix} \sum_{j} {b_j \choose 2} \end{matrix}
    ]/{N \choose 2}}
\end{equation}
where $n_{ij}$ denotes the number of cells in both cluster $i$ of $U$ and cluster $j$ of $V$, and $a_i$ denotes the count of cells assigned to cluster $i$ of $U$, $b_j$ indicates the count of cells assigned to cluster $j$ of $V$. 

\subsection{The CICL Method}
\subsubsection{Overview}
CICL is a cluster-aware iterative contrastive learning method designed for clustering scRNA-seq data, its framework is illustrated in Fig.~\ref{fig:workflow}. Specifically, in the model training phase, we first generate two augmented views $X_{aug1}$ and $X_{aug2}$ of the raw data $X$ by adding noise that is randomly sampled from Gaussian $N(0, 1)$ and is mapped to the range [0, 1] via a linear transformation.
Then, $X$, $X_{aug1}$ and $X_{aug2}$ are input into a transformer encoder~\citep{vaswani2017attention} to obtain their representations $H$, $H_{aug1}$ and $H_{aug2}$, respectively. 
Next, we perform K-means on $H$ to get the centroid matrix $C = [c_{1}, c_{2}, ..., c_{K}]$ where $c_{i}$ is the centroid vector of cluster $i$. The number of centroids is equal to the number of cell subtypes (or clusters) in the training dataset. 
After that, $H$ and $C$ are fed to the clustering-head and generate a pseudo-label for each cell.  
Meanwhile, the projection-head encodes $H_{aug1}$ and $H_{aug2}$ to obtain their projections $Z_{aug1}$ and $Z_{aug2}$. 
Finally, in addition to the traditional instance-wise contrastive loss, we propose a novel cluster-aware contrastive loss to align the positive pairs and contrast the negative pairs simultaneously, which takes the projections $Z_{aug1}$, $Z_{aug2}$ and pseudo-labels as input. We construct the positive pairs by an instance-wise way and a pseudo-label based way. In particular, an instance-wise positive pair consists of the representations of the two augmented copies of each cell, and a pseudo-label positive pair is formed by the representations of two augmented copies of two cells with similar pseudo-label (i.e., belonging to the same cluster). This training process goes iteratively. 
In the clustering phase, the input data $X$ are preprocessed and encoded by the trained transformer encoder to obtain the representation $H$. Then, $H$ are clustered by K-means to generate the final clustering result.
In the following sections, we present the major components of our method in detail.

\subsubsection{Transformer Encoder}
The raw scRNA-seq data is modeled as a matrix $X \in \mathbb{R}^{N \times G}$ where $N$ indicates the number of cells and  $G$ denotes the number of genes. 
To begin with, we construct augmented data by adding Gaussian noise, and two augmented copies (or views) $X_{aug1}$ and $X_{aug2}$ are generated for $X$.
Then, we encode $X_{aug1}$, $X_{aug2}$ and $X$ by a transformer encoder, which has four layers, each of which consists of two networks: a multi-head self-attention network and a position-wise fully connected feed-forward network, each of them is followed by a residual connection and layer normalization.
For example, give the input $X$ of the self-attention layer, the output is as follows: 
\begin{equation}
    H_{MulitHead} = Concat(Att_{1}(XW_{1}^{v}), ..., Att_{h}(XW_{h}^{v}))W^{O}
\end{equation}
where $W_{i}^{v} \in \mathbb{R}^{G \times d} $ and $W^{O} \in \mathbb{R}^{hd \times G}$ are learnable parameter matrices, $h$ is the number of heads. And $Att_{i}$ is evaluated by 
\begin{equation}
    Att_{i} = softmax(\frac{XW_{i}^{q} \times (XW_{i}^{k})^{\mathsf{T}}}{\sqrt{d}}) , i = 1, 2, ..., h 
\end{equation}
Above, $W_{i}^{q} \in \mathbb{R}^{G \times d} $ and $W_{i}^{k} \in \mathbb{R}^{G \times d}$ are learnable parameter matrices.
Then, after the residual connection and layer normalization, we have
\begin{equation}
    H_{res} = LayerNorm(X + H_{MulitHead})
\end{equation}

The fully connected feed-forward network consists of two linear layers, following a rectified linear activation function~(ReLU), so we have
\begin{equation}
    H_{fc} = ReLU(H_{res}W_{1})W_{2}
\end{equation}
where $W_{1}$ and $W_{2}$ are learnable parameter matrices. Finally, the output of the $i$-th layer of
the transformer encoder is
\begin{equation}
    H_{i} = LayerNorm(H_{fc} + H_{res})
\end{equation}


\subsubsection{Clustering-head and Pseudo-label Generation}
Traditional contrastive loss suffers from sampling bias~\citep{oord2018representation}. For example, given a cell $i$, all the other cells are considered as its negative samples. However, these negative samples contain some cells of the same type as cell $i$, which will be undesirably pushed away from cell $i$ in the representation space by current contrastive learning. 

To address this problem, CICL employs a cluster-aware contrastive learning strategy. To this end, we cluster the training data $H$ by K-means, and each cluster is characterized by its centroid $c_{i}$ in the representation space, which will be updated iteratively.
Then, in the \emph{clustering-head}, we use the Student’s t-distribution to compute the probability $q_{ij}$ that cell $i$ belongs to the $j$-th cluster,  
\begin{equation}\label{alpha}
    q_{ij} = \frac{(1+ \| h_{i} - c_{j}) \|_{2}^{2} / \alpha)^{-\frac{\alpha + 1}{2}}}{\sum_{k = 1}^{K}(1+ \| h_{i} - c_{k}) \|_{2}^{2} / \alpha)^{-\frac{\alpha + 1}{2}}}
\end{equation}
where $h_{i}$ is the representation of cell $i$ in $H$. $\alpha$ is the degree of freedom of the Student’s t-distribution, we set $\alpha = 1$ in this paper. Finally, we obtain the pseudo-label $l_{i}$ of cell $i$ by the probability vector $q_{i}$=$(q_{i1},q_{i2},...,q_{iK})$ as follows: 
\begin{equation}\label{label assign}
    l_{i} = label\_assign(q_i)
\end{equation}
where $label\_assign$ is a function that returns the cluster index corresponding the maximum $q_{ij}$ ($j \in 1, 2, ..., K$). Thus, we obtain the pseudo-labels $L$=($l_1$, $l_2$, ..., $l_N$) of all cells, which are used for contrastive loss computation. Note that each cell and its two augmented copies have the same pseudo-label. We use the term ``pseudo-label''  because they are just intermediate (not final) cluster labels. 

\subsubsection{Projection-head and Contrastive Learning Losses}
We project $H_{aug1}$ and $H_{aug2}$ to obtain projections $Z_{aug1}$ and $Z_{aug2}$ by the projection-head, which is composed of a two-layer perceptron. Formally,
\begin{equation}
    Z_{aug1} = W_{3}ReLU(W_{4}H_{aug1})
\end{equation}
\begin{equation}
    Z_{aug2} = W_{3}ReLU(W_{4}H_{aug2})
\end{equation}
where $W_{3}$ and $W_{4}$ are learnable parameters. ReLU is the activation function.
Let $z_{i}$ and $z^{'}_{i}$ be the $i$-th row of $Z_{aug1}$ and $Z_{aug2}$ respectively, which correspond to the representations of cell $i$ in the two augmented views. 
For $z_{i}$, we not only treat $z_{i}$ and $z^{'}_{i}$, but also $z_{i}$ and any other sample of the same cluster in terms of pseudo-label as a positive pair, while $z_{i}$ and any sample of the other clusters as a negative pair. Given the batch size $B$, we consider two losses as follows: 

\textbf{Instance-wise contrastive loss.} CICL computes the infoNCE loss~\citep{chen2020simple} for each cell. For cell $i$ with two views $z_{i}$ and $z_{i}^{'}$, its contrastive loss in terms of $z_{i}$ is
\begin{equation}
    \resizebox{1\hsize}{!}{$
    l_{ins}(z_{i}) = -log \frac{exp(sim{(z_{i}, z_{i}^{'})}/T)}{\sum_{m=1}^{B}\mathbb{I}_{i \neq m} exp(sim{(z_{i}, z_{m})}/T) + \sum_{m=1}^{B} exp(sim{(z_{i}, z_{m}^{'})}/T)}
    $}
\end{equation}
where $\mathbb{I}_{i \neq m}$ is an indicator function whose value is 1 if $i \neq m$, otherwise is 0. $T$ is the temperature parameter set to 0.5 in our paper.
The similarity function $sim(.,.)$ adopts point product or cosine similarity, i.e.,
\begin{equation}
    sim{(z_{i}, z_{i}^{'})} = \frac{\|z_{i}^{T}z_{i}^{'}\|}{\|z_{i}\| \|z_{i}^{'}\|}
\end{equation}

The overall instance-wise contrastive loss is
\begin{equation}\label{ins loss}
    \mathcal{L}_{ins} = \frac{1}{2B}\sum^{B}_{i=1}[l_{ins}(z_{i}) + l_{ins}(z^{'}_{i})]
\end{equation}
where $l_{ins}(z^{'}_{i})$ is cell $i$'s contrastive loss in terms of $z^{'}_{i}$. With this instance-wise contrastive loss, CICL can learn the representations well by pulling positive pairs together and pushing negative pairs away in the cell representation space.

\textbf{Cluster-aware contrastive loss.} To solve the sample bias, we propose a novel cluster-aware contrastive loss, which is evaluated with the pseudo-labels $L$, $Z_{aug1}$ and $Z_{aug2}$. We treat the pairs of representations of the same pseudo-label as positive pairs, and the remaining pairs as negative pairs. For cell $i$, its cluster-aware contrastive 
 loss in terms of $z_{i}$ is as follows:
\begin{equation}
  \resizebox{1.05\hsize}{!}{$
    l_{clu}(z_{i}) = -log \frac{\sum_{j}^{B} E_{z_{i}, z^{'}_{j} \in l_{i}} \cdot exp(sim{(z_{i}, z^{'}_{j})}/T) + \sum_{j}^{B} E_{z_{i}, z_{j} \in l_{i}} \cdot \mathbb{I}_{i \neq j} \cdot exp(sim{(z_{i}, z_{j})}/T)}{\sum_{j}^{B} E_{z_{i}, z^{'}_{j} \notin l_{i}} \cdot exp(sim{(z_{i}, z^{'}_{j})}/T) + \sum_{j}^{B} E_{z_{i}, z_{j} \notin l_{i}} \cdot exp(sim{(z_{i}, z_{j})}/T)}
    $}
\end{equation}
where $l_{i}$ is the pseudo-label of $z_{i}$.
$E_{z_{i}, z^{'}_{j} \in l_{i}}$ is an indicator function whose value is 1 if the label of $z^{'}_{j}$ is $l_{i}$, and 0 otherwise. $E_{z_{i}, z^{'}_{j} \notin l_{i}}$ is also an indicator function whose value is 0 if the label of $z^{'}_{j}$ is $l_{i}$, and 1 otherwise.
The overall cluster-aware loss is as follows:
\begin{equation}\label{clu loss}
    \mathcal{L}_{clu} = \frac{1}{2B}\sum_{i=1}^{B}[l_{clu}(z_{i}) + l_{clu}(z^{'}_{i})]
\end{equation}
where $l_{clu}(z^{'}_{i})$ is cell $i$'s cluster-aware contrastive 
loss in terms of $z^{'}_{i}$. This loss is particularly effective because it tries to minimize the distance between cells of similar cluster and maximize the distance between cells of different clusters.

Finally, by combining $\mathcal{L}_{ins}$ and $\mathcal{L}_{clu}$ with the hyperparameter $\lambda$, we have the whole loss function as follows:
\begin{equation}\label{equ:loss}
    \mathcal{L} = \mathcal{L}_{ins} + \lambda\mathcal{L}_{clu}
\end{equation}
We set $\lambda = 0.1$ in our experiments. A small $\lambda$ can prevent the negative effect of clustering error of K-means.
With this loss, CICL exploits the cluster structure underlying the data to achieve simultaneous optimization of data representation and cluster label assignment. 
Compared with traditional contrastive learning, ours is iterative contrastive learning, which iteratively learns the representations of cells in the direction favorable for clustering. 

\subsection{Algorithm}
Here, we present the algorithm of our method in Alg.~\ref{CICL}, which consists of two phases: the training phase and the clustering phase. In the training phase, 
in each epoch we first randomly split the training data $X^{train}$ into $n_B$=$[X^{train}/S_B]$ minbatches. And for each minbatch $X^j$, we 
generate two augmented views $X^j_{aug1}$ and $X^j_{aug2}$.  Next, we obtain the representations $H^j_{aug1}$, $H^j_{aug2}$ and $H^j$ of $X^j_{aug1}$, $X^j_{aug2}$ and $X^j$ by a transformer encoder. We perform K-means on $H^j$ to get centroid matrix $C^j$, and then the pseudo-labels $L^j$ are obtained from $H^j$ and $C^j$. 
Meanwhile, $H^j_{aug1}$, $H^j_{aug2}$ are input into the projection-head for obtaining projections $Z^j_{aug1}$ and $Z^j_{aug2}$. The whole loss $\mathcal{L}$ consists of $\mathcal{L}_{ins}$ and $\mathcal{L}_{clu}$, which are computed with $Z^j_{aug1}$,  $Z^j_{aug2}$ and $L^j$ by Equ.~(\ref{ins loss}) and Equ.~(\ref{clu loss}) respectively. In the training phase, we use K-means to cluster the representations $H^{test}$ of testing data $X^{test}$, encoded by the trained transformer encoder, to generate the clustering result $R^{test}$. 

\IncMargin{1em}
\begin{algorithm}[h]
  \SetKwData{Left}{left}\SetKwData{This}{this}\SetKwData{Up}{up}
  \SetKwFunction{Union}{Union}\SetKwFunction{FindCompress}{FindCompress}
  \SetKwInOut{Input}{Input}\SetKwInOut{Output}{Output}

  \Input{Training data $X^{train}$;\\Testing data $X^{test}$;\\The number of training epochs $n_{epoch}$;\\Minibatch size $S_B$;}
  \Output{Clustering result $R^{test}$ of $X^{test}$;}
  \BlankLine

  {\textbf{Training phase:}}\\
  \For{$i$=1 \KwTo $n_{epoch}$}{{$n_B$=$[X^{train}/S_B]$};\\
    Randomly split $X^{train}$ into $n_B$ minibatches;\\ 
    \For{each minibatch $X^j$}{\label{forins}
        {Generate views $X^j_{aug1}$ and $X^j_{aug2}$ of $X^j$ by adding Gaussian noise;} \\
        Obtain representations $H^j$, $H^j_{aug1}$, $H^j_{aug2}$ of $X^j$, $X^j_{aug1}$, $X^j_{aug2}$ by a transformer encoder; \\
        Generate centroid matrix $C^j$ of $H^j$ by K-means;\\
        Generate pseudo-labels $L^j$ with $H^j$ and $C^j$ by Equ.~\ref{alpha} and Equ.~\ref{label assign}; \\
        Obtain projections $Z^j_{aug1}$ and $Z^j_{aug2}$ of $H^j_{aug1}$ and $H^j_{aug2}$ by the projection-head; \\
        Compute $\mathcal{L}_{ins}$ using $Z^j_{aug1}$ and $Z^j_{aug2}$ by Equ.~\ref{ins loss}; \\
        Compute $\mathcal{L}_{clu}$ using $Z^j_{aug1}$, $Z^j_{aug2}$ and $L$ by Equ.~\ref{clu loss}; \\
        Compute the whole loss $\mathcal{L} = \mathcal{L}_{ins} + \lambda \mathcal{L}_{clu}$ with $\lambda = 0.1$;\\
        Update the parameters of the model according to the loss $\mathcal{L}$.
    }
    
  }
  \BlankLine
 
  {\textbf{Clustering phase:}}\\
  {Obtain the representations $H^{test}$ of $X^{test}$ by the trained transformer encoder;}\\
  {Cluster representations $H^{test}$ by K-means and get the clustering result $R^{test}$.}
  \caption{CICL~(Cluster-aware Iterative Contrastive Learning for scRNA-seq data clustering)}\label{CICL}
  
\end{algorithm}\DecMargin{1em}

\section{Experiments and Results}

\subsection{Implementation Details and Experimental Setup}
Here, we present the implementation details of our method and the experimental setup in Table~\ref{hyperp}. CICL uses similar parameters on all of the datasets in Table~\ref{dataset}, and all compared methods use default parameters provided in their original papers. All the experiments in this paper are conducted on 4 NVIDIA RTX3090 GPUs.

\begin{table}[h]    
    \centering
    \vspace{-0.5cm}
    \fontsize{9}{9}\selectfont  
    \caption{Implementation details and experimental setup.}
    \begin{tabular}{|c|cc|c|}
     \Xhline{1pt}
         Category& \multicolumn{2}{c|}{Parameter}&Value \\
        \Xhline{0.5pt}
        \multirow{3}*{Model architecture}
        & \multicolumn{2}{c|}{The number of Transformer encoder layers} & 4 \\
        & \multicolumn{2}{c|}{The feed-forward network of Transformer encoder layers} & [1024, 1024] \\
        & \multicolumn{2}{c|}{The projection-head} & [1024, 512]\\
        
        \Xhline{1pt}
        
       \multirow{6}*{Experimental setup}
        & \multicolumn{2}{c|}{Learning rate} &\num{e-5}\\
        & \multicolumn{2}{c|}{Batch size } & 6000 \\
        & \multicolumn{2}{c|}{Optimizer} & Adam \\
        & \multicolumn{2}{c|}{The number of training epochs} & 1000 \\
        & \multicolumn{2}{c|}{$\lambda$ (in Equ.~(\ref{equ:loss}) of the paper)} & 0.1 \\
        & \multicolumn{2}{c|}{$\alpha$ (in Equ.~(\ref{alpha}) of the paper)} & 1 \\
        
        \Xhline{1pt}
    \end{tabular}
    \label{hyperp}
    \vspace{-0.7cm}
\end{table}

\subsection{Compared Existing Methods}
We compare CICL with 8 existing scRNA-seq data clustering methods, including a graph-based method Seurat~\citep{satija2015spatial}, a multi-kernel learning method SIMLR~\citep{wang2018simlr}, a transfer learning method ItClust~\citep{hu2020iterative}, a contrastive learning method CLEAR~\citep{han2022self}, a deep graph embedding based method GraphSCC~\citep{zeng2020accurately}, and three deep embedding based methods scDeepCluster~\citep{tian2019clustering}, scDHA~\citep{tran2021fast} and scVI~\citep{lopez2018deep}. More information of these methods is as follows:
\begin{itemize}
    \item \textbf{Seurat}~\citep{satija2015spatial} is a widely used pipeline for single-cell gene expression data analysis. It performs dimension reduction first, then employs Louvain method on the shared nearest neighbor graph.
    \item \textbf{SIMLR}~\citep{wang2018simlr} combines multiple cores to learn the similarity between samples and performs spectral clustering.
    \item \textbf{ItClust}~\citep{hu2020iterative} trains a neural network to extract information from a well-labeled source dataset, then initializes the target network with parameters estimated from the training network.
     \item \textbf{CLEAR}~\citep{han2022self} is a self-supervised contrastive learning-based integrative scRNA-seq data analysis tool. It introduces a novel data augmentation method and performs contrastive learning by InfoNCE loss.
    \item \textbf{GraphSCC}~\citep{zeng2020accurately} extracts the structural relationships between cells using a graph convolutional network, and optimizes the representations by a dual self-supervised module.
    \item \textbf{scDeepCluster}~\citep{tian2019clustering} adds a ZINB distribution model simulating the distribution of scRNA-seq data to the denoising autoencoder, and learns feature representations and clusters by explicit modeling of scRNA-seq data. 
    \item \textbf{scDHA}~\citep{tran2021fast} first exploits a non-negative kernel autoencoder to do dimension reduction and then projects the data onto a low-dimensional space with a self-learning network based on variational autoencoder~(VAE).
    \item \textbf{scVI}~\citep{lopez2018deep} is a comprehensive tool for the analysis of scRNA-seq data. It models scRNA-seq data in a deep generative manner with the ZINB model and variational autoencoder.
\end{itemize}

\subsection{Performance Comparison}
Table \ref{tresults} summarizes the clustering performance of CICL and 8 existing methods on 25 scRNA-seq datasets. 
CICL achieves the best ARI and NMI on 10 and 9 datasets, and the 2nd best ARI and NMI on 7 and 10 datasets, respectively. On average, our method obtains the best ARI (0.7757) and NMI (0.8057) on the 25 datasets. In particular, CICL surpasses scDHA by 13.87\% and 4.96\% in terms of ARI and NMI on average, which shows the outstanding clustering performance of our method.  We can also see that CICL performs excellently on large datasets such as Bach~(23184 cells), havatin~(48266 cells), QX\_Trachea~(11269 cells), QX\_Spleen~(9552 cells) and Wang\_Lung~(9519 cells). Furthermore, our method also achieves good clustering scores on datasets with more than 10 subtypes of cells, such as muraro~(10 subtypes), pollen~(11 subtypes), QS\_Lung~(11 subtypes) and Young~(11 subtypes). In summary, CICL surpasses the existing methods by from 14\% to 280\%, and from 5\% to 133\% on average in terms of performance metrics ARI and NMI,  respectively.

\begin{table}[ht]
    \centering
    \vspace{-0.5cm}
    \fontsize{5}{8}\selectfont
    \caption{Performance comparison with existing methods. The best results are \textbf{boldfaced} and the 2nd best results are \underline{underlined}. `$\backslash$' means data unavailable.}
    \begin{tabular}{c|cc|c|cccccccc}
     \Xhline{1pt}
         Metric& \multicolumn{2}{c|}{Datasets}& CICL~(\textbf{ours})&CLEAR&GraphSCC&scDeepCluster&scDHA&scVI&Seurat&SIMLR&ItClust  \\
        \Xhline{0.5pt}
        \multirow{26}*{ARI}& \multicolumn{2}{c|}{Adam} &\underline{0.7175}&0.5247&0.2444&0.6690&0.3924&\textbf{0.8519}&0.6659&0.3454&0.2640 \\
        & \multicolumn{2}{c|}{Bach} &\textbf{0.8071}&0.5776&0.6630&\underline{0.7874}&0.6918&0.6595&0.5093&0.6690&0.2593 \\
        & \multicolumn{2}{c|}{Bladder} &\textbf{0.9885}&\underline{0.9852}&\underline{0.9852}&0.7076&0.9803&0.9803&0.3401&0.9755&0.2330 \\
        & \multicolumn{2}{c|}{deng} &\textbf{0.8982}&0.6260&0.5531&0.5611&0.5634&0.4207&0.4142&\underline{0.7741}&0.0596 \\
        & \multicolumn{2}{c|}{Diaphragm} &\underline{0.9309}&0.9075&0.9162&0.3841&\textbf{0.9387}&0.4789&0.4988&0.4552&0.1074 \\
        & \multicolumn{2}{c|}{hrvatin} &0.8730&0.6669&\underline{0.8797}&0.8646&0.8715&\textbf{0.8876}&0.6585& $\backslash$&0.3401 \\
       &  \multicolumn{2}{c|}{Klein} &0.8180&0.5392&0.2044&\textbf{0.8665}&0.6653&0.7765&\underline{0.8322}&0.6143&0.1152 \\
        & \multicolumn{2}{c|}{kolodziejczyk} &\underline{0.6150}&0.4261&0.4858&0.0590&\textbf{0.9960}&0.5695&0.5174&0.5173&$\backslash$ \\
         &\multicolumn{2}{c|}{Limb\_Muscle} &0.8961&0.6565&\textbf{0.9243}&0.2570&\underline{0.9138}&0.5126&0.4577&0.5090&0.1196 \\
        & \multicolumn{2}{c|}{Mammary\_Gland} &\underline{0.9226}&0.8993&0.8921&0.3837&\textbf{0.9259}&0.5364&0.4429&0.5585&0.0493 \\
        & \multicolumn{2}{c|}{muraro} &\textbf{0.9383}&0.5134&{0.8761}&\underline{0.9204}&0.6580&0.5809&0.6705&0.6082&$\backslash$ \\
        & \multicolumn{2}{c|}{Plasschaert} &0.4278&0.4052&0.2203&\underline{0.5157}&0.3945&0.3836&0.4883&\textbf{0.7394}&0.1125 \\
       &  \multicolumn{2}{c|}{pollen} &\underline{0.9117}&0.7500&0.5955&0.4849&\textbf{0.9529}&0.8597&0.7280&0.7539&0.2992 \\
        & \multicolumn{2}{c|}{Qx\_Spleen} &\underline{0.4923}&0.2838&0.0764&\textbf{0.6026}&0.4850&0.3134&0.2476&0.3889&0.1199 \\
        & \multicolumn{2}{c|}{Qx\_Trachea} &0.4949&0.2868&0.1437&0.4767&\underline{0.4979}&\textbf{0.5261}&0.1829&0.3647&0.2229 \\
        & \multicolumn{2}{c|}{QS\_Diaphragm} &0.9561&\underline{0.9571}&0.4251&0.6829&\textbf{0.9785}&0.6603&0.6872&0.6559&0.1948 \\
        & \multicolumn{2}{c|}{QS\_Heart} &\textbf{0.9660}&0.5386&0.3016&0.6146&0.6322&0.6136&0.4207&\underline{0.8758}&0.4113 \\
        & \multicolumn{2}{c|}{QS\_Limb\_Muscle} &0.9094&0.8918&0.5303&0.5966&\underline{0.9138}&0.6555&0.5955&\textbf{0.9307}&0.2916 \\
        & \multicolumn{2}{c|}{QS\_Lung} &\textbf{0.7297}&0.5773&0.4659&0.4653&$\backslash$&0.3832&0.5034&\underline{0.7034}&0.2350 \\
       &  \multicolumn{2}{c|}{QS\_Trachea} &\textbf{0.5841}&0.5445&\underline{0.5670}&0.4276&0.4903&0.5513&0.2552&0.4753&0.2547 \\
        & \multicolumn{2}{c|}{Romanov} &\textbf{0.6886}&\underline{0.5920}&0.3181&0.5827&0.5388&0.4902&0.4460&0.5807&0.2682 \\
        & \multicolumn{2}{c|}{Tosches\_turtle} &0.4228&0.3432&0.4188&\underline{0.6016}&0.4539&\textbf{0.6045}&0.5919&0.5208&0.1302 \\
       &  \multicolumn{2}{c|}{Wang\_lung} &\textbf{0.9469}&0.0549&0.7358&0.7259&\underline{0.7824}&0.0701&0.2183&-0.0079&0.0488 \\
       &  \multicolumn{2}{c|}{yan} &\underline{0.7989}&0.6584&0.5743&\textbf{0.8029}&0.6891&0.5671&0.6565&0.6117&0.3118 \\
        & \multicolumn{2}{c|}{Young} &\textbf{0.6578}&0.3806&0.2133&0.5303&0.4077&0.5444&\underline{0.5965}&0.2688&0.2520 \\
        \Xcline{2-12}{0.5pt}
        & \multicolumn{2}{c|}{AVERAGE} &\textbf{0.7757}&0.5835&0.5284&0.5828&\underline{0.6812}&0.5791&0.5050&0.5787&0.2044 \\
        \Xhline{1pt}

       \multirow{25}*{NMI}& \multicolumn{2}{c|}{Adam} &0.7681&0.6491&0.5606&0.7565&0.5646&\textbf{0.8436}&\underline{0.7936}&0.5473&0.4241 \\
        & \multicolumn{2}{c|}{Bach} &\textbf{0.8086}&0.7620&0.7130&\underline{0.7985}&0.7901&0.7730&0.7374&0.7906&0.3876 \\
        & \multicolumn{2}{c|}{Bladder} &\textbf{0.9709}&\underline{0.9679}&\underline{0.9679}&0.5941&0.9593&0.9593&0.5122&0.9449&0.3179 \\
        & \multicolumn{2}{c|}{deng} &\textbf{0.8469}&\underline{0.8056}&0.7204&0.7717&0.8023&0.6951&0.6654&0.7982&0.2612 \\
        & \multicolumn{2}{c|}{Diaphragm} &\textbf{0.8660}&0.8139&0.8366&0.3577&\underline{0.8585}&0.6428&0.6056&0.5915&0.2013 \\
        & \multicolumn{2}{c|}{hrvatin} &\underline{0.9085}&0.7816&\textbf{0.9122}&0.8628&0.9034&0.9018&0.8257&$\backslash$&0.5135 \\
       &  \multicolumn{2}{c|}{Klein} &0.8100&0.6781&0.2434&\textbf{0.9025}&0.6926&0.8197&\underline{0.8545}&0.6205&0.2107 \\
        & \multicolumn{2}{c|}{kolodziejczyk} &0.6813&0.3863&0.4561&0.0584&\textbf{0.9915}&0.5756&\underline{0.7184}&0.6722&$\backslash$ \\
         &\multicolumn{2}{c|}{Limb\_Muscle} &0.8212&0.7492&\textbf{0.8587}&0.1703&\underline{0.8401}&0.6648&0.5318&0.6534&0.2340 \\
        & \multicolumn{2}{c|}{Mammary\_Gland} &\underline{0.8859}&0.8485&0.8536&0.3271&\textbf{0.8891}&0.7022&0.5404&0.7606&0.2019 \\
        & \multicolumn{2}{c|}{muraro} &\textbf{0.8846}&0.7014&{0.8328}&\underline{0.8779}&0.7908&0.7770&0.8150&0.7746&$\backslash$ \\
        & \multicolumn{2}{c|}{Plasschaert} &0.6164&0.6041&0.3868&\underline{0.6745}&0.6093&0.6095&0.6665&\textbf{0.7610}&0.1719 \\
       &  \multicolumn{2}{c|}{pollen} &\underline{0.9323}&0.8565&0.7868&0.6142&\textbf{0.9446}&0.9174&0.8897&0.8705&0.4984 \\
        & \multicolumn{2}{c|}{Qx\_Spleen} &\textbf{0.6585}&0.4950&0.1344&0.6005&\underline{0.6221}&0.5186&0.5137&0.5048&0.2542 \\
        & \multicolumn{2}{c|}{Qx\_Trachea} &\underline{0.7124}&0.5449&0.1190&0.6610&0.6525&\textbf{0.7200}&0.5358&0.4976&0.4783 \\
        & \multicolumn{2}{c|}{QS\_Diaphragm} &\underline{0.9412}&0.9396&0.4146&0.7508&\textbf{0.9640}&0.8112&0.8383&0.7974&0.3899 \\
        & \multicolumn{2}{c|}{QS\_Heart} &\textbf{0.9328}&0.7254&0.4796&0.7022&0.7822&0.7702&0.7233&\underline{0.7832}&0.4990 \\
        & \multicolumn{2}{c|}{QS\_Limb\_Muscle} &\underline{0.8810}&0.8500&0.7009&0.6345&0.8184&0.8185&0.8124&\textbf{0.8839}&0.4800 \\
        & \multicolumn{2}{c|}{QS\_Lung} &\underline{0.7786}&0.7214&0.5985&0.6885&$\backslash$&0.6934&\textbf{0.7876}&0.7575&0.4434 \\
       &  \multicolumn{2}{c|}{QS\_Trachea} &\textbf{0.6683}&0.6585&0.5406&0.5531&0.5976&\underline{0.6639}&0.6066&0.6075&0.4193 \\
        & \multicolumn{2}{c|}{Romanov} &\underline{0.6380}&0.5820&0.4792&0.5865&0.5847&0.5522&\textbf{0.6495}&0.5681&0.3565 \\
        & \multicolumn{2}{c|}{Tosches\_turtle} &0.6739&0.6287&0.4973&0.7544&0.6698&\underline{0.7589}&\textbf{0.8087}&0.6306&0.3106 \\
       &  \multicolumn{2}{c|}{Wang\_lung} &\textbf{0.8778}&0.2237&0.6581&0.6494&\underline{0.7000}&0.2255&0.4259&0.0050&0.0461 \\
       &  \multicolumn{2}{c|}{yan} &\underline{0.8307}&0.7951&0.7322&\textbf{0.8566}&0.8269&0.7418&0.7476&0.7695&0.4969 \\
        & \multicolumn{2}{c|}{Young} &\underline{0.7479}&0.5808&0.2102&0.6973&0.5671&0.6553&\textbf{0.7619}&0.4373&0.3642    \\
        \Xcline{2-12}{0.5pt}
        & \multicolumn{2}{c|}{AVERAGE} &\textbf{0.8057}&0.6940&0.5877&0.6360&\underline{0.7676}&0.7125&0.6947&0.6678&0.3461 \\
        \Xhline{1pt}
    \end{tabular}
    \label{tresults}
    \vspace{-0.7cm}
\end{table}

Note that the latest contrastive learning-based method CLEAR does not show advantages over the other methods on the 25 datasets. However, our method achieves excellent results, thanks to the proposed cluster-aware iterative contrastive learning mechanism. 

\subsection{Visualization with Low-dimensional Representations}
\begin{figure*}[htbp]
	\centering
    \vspace{-0.4cm}
	\subfloat{\includegraphics[width=0.9\textwidth]{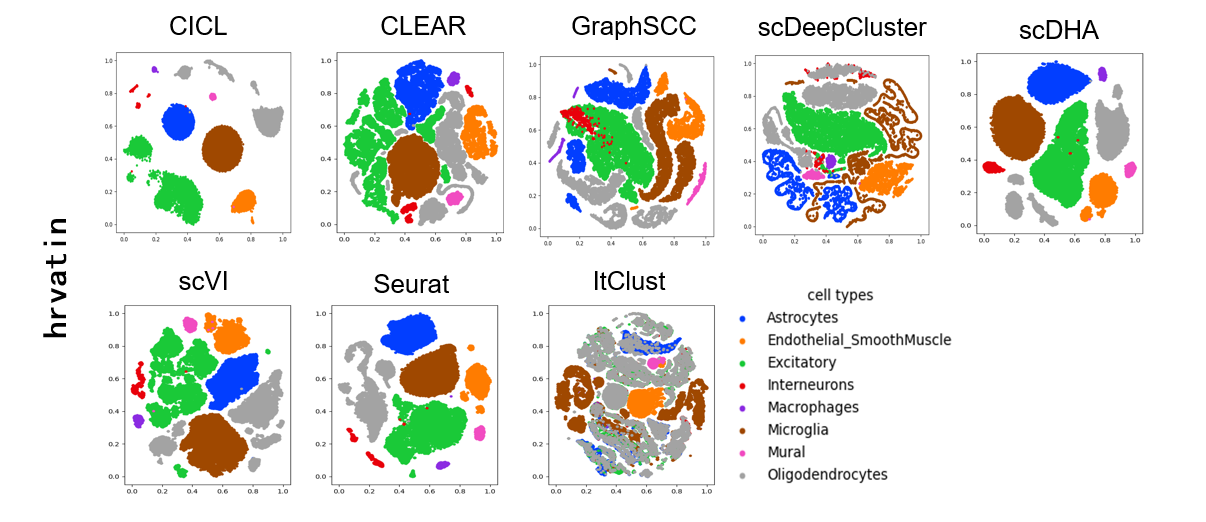}} \\
    \vspace{-0.3cm}
	\subfloat{\includegraphics[width=0.9\textwidth]{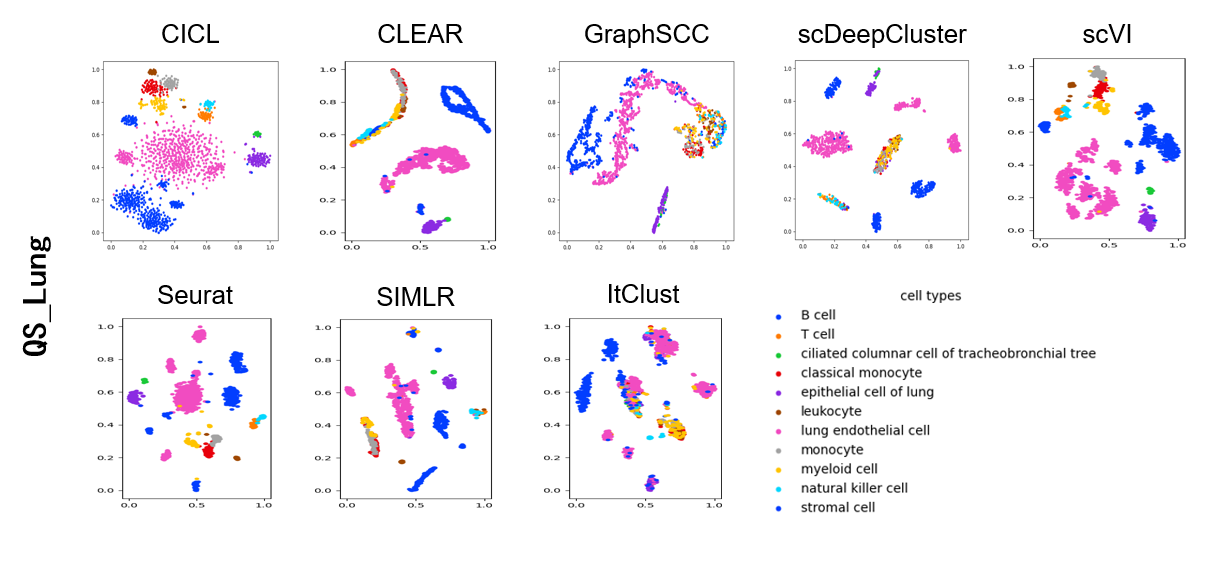}} \\
    \vspace{-0.3cm}
	\subfloat{\includegraphics[width=0.9\textwidth]{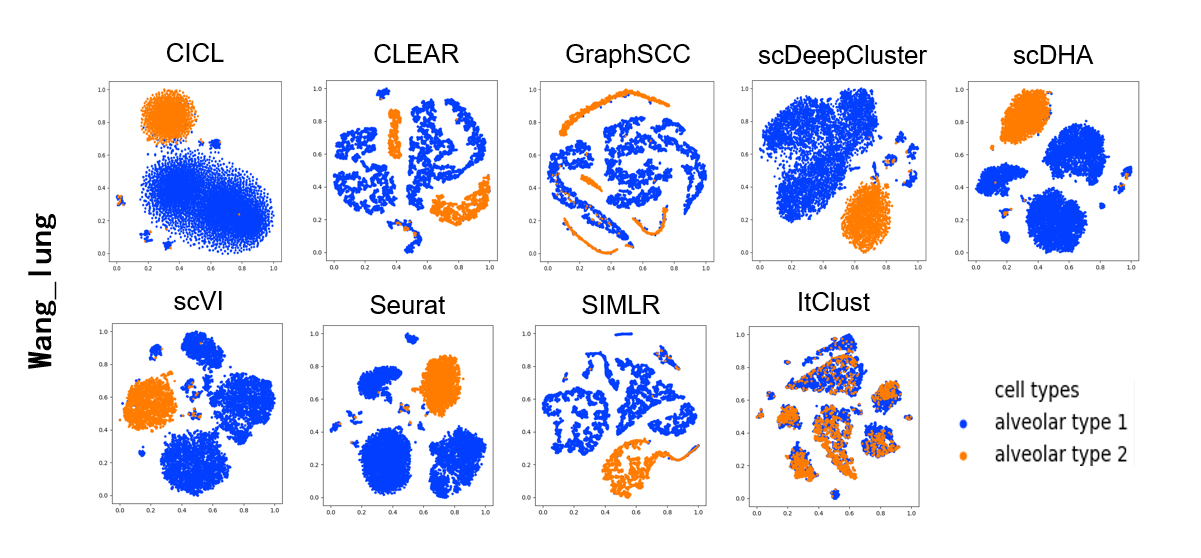}}
    \vspace{-0.3cm}
	\caption{Visualization of clustering results on the hrvatin, QS\_Lung and Wang\_lung datasets by t-SNE. 
 }
    \label{fig:vis}
    \vspace{-0.3cm}
\end{figure*}

In cellular heterogeneity analysis, visualization is an intuitive and effective way to display different cell types. We use t-SNE~\citep{van2008visualizing} to project the representations of cells into a two-dimensional space and visualize them in Fig.~\ref{fig:vis}. As we can see, CICL learns to embed cells of the same type within the same cluster while separating cells of different types well into different clusters, producing similar clustering results to the ground truth cell annotations.
The clustering result of CICL is superior to that of the other methods on the hrvatin dataset. Although the performance of scDHA and scVI is also good, they divide the oligodendrocyte cells into multiple clusters. 
Furthermore, CICL performs well on QS\_Lung, the cells of different types are effectively separated in the embedding space, which is much better than with the other methods.
As for the Wang\_Lung dataset with two subtypes, CICL not only achieves the best ARI (see Table \ref{tresults}) but also exhibits the best clustering visualization effect: the data is grouped into two distinct clusters.

\begin{figure}[htbp]
  \centering
  \vspace{-0.3cm}
  \includegraphics[width=\textwidth]{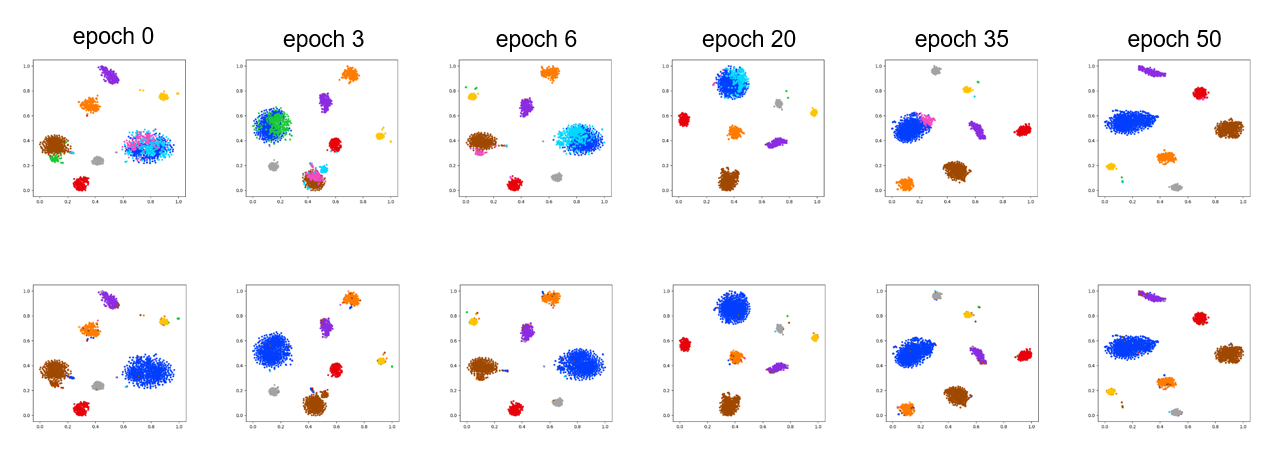}
  \vspace{-0.5cm}
  \caption[muraro dataset]{Iterative clustering results on the muraro dataset, which show the visualization of embedded cells in the learning process. Upper:   clustering results of our method; Lower: ground truth. Each color indicates a type of cells.}
  
  \label{fig:cic}
  \vspace{-0.1cm}
\end{figure}

Fig.~\ref{fig:cic} illustrates the clustering process of our iterative contrastive learning on the muraro dataset. The upper and lower figures represent the clustering results of our method and the ground truth, respectively.
We can see that various types of cells are distributed chaotically in the early epochs~(e.g. epoch = 0, 3, 6). 
However, as the iterative learning goes, CICL split different types of cells with growing accuracy. At epoch 50, our method correctly clusters the data. In summary, 
CICL is able to gradually refine the clustering outcome, and eventually makes the clustering result match with the ground truth. This demonstrates that our model iteratively learns more and more accurate cell representations.  

\begin{figure}[htbp]
	\vspace{-0.6cm}
        \centering
	\subfloat{\includegraphics[width=0.4\textwidth]{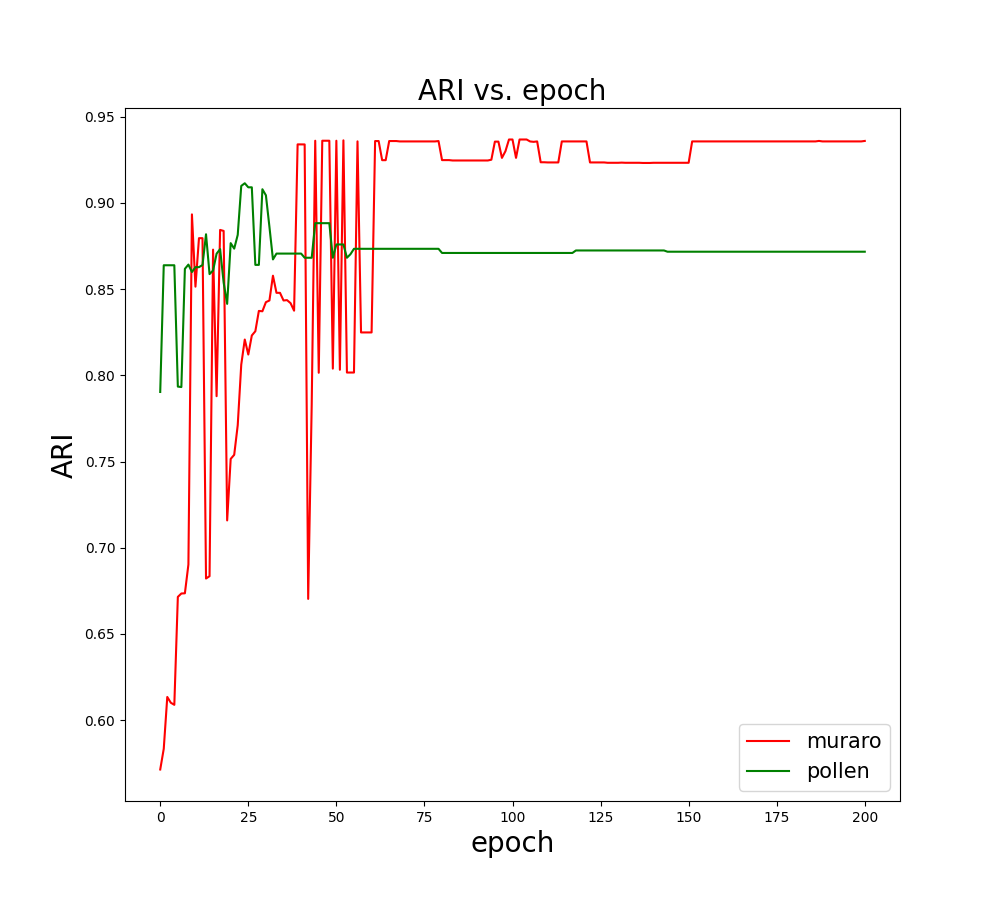}} 
	\subfloat{\includegraphics[width=0.4\textwidth]{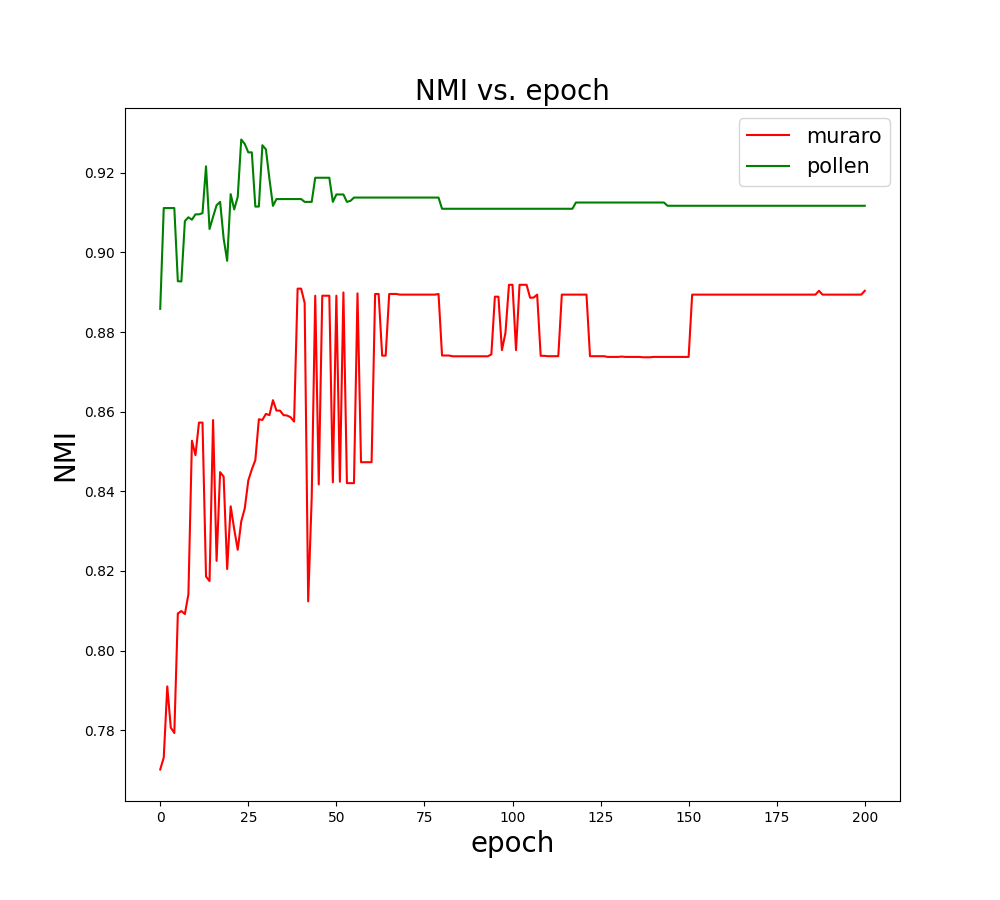}}
        \vspace{-0.4cm}
	\caption{The ARI and NMI results during the iterative learning process on the muraro and pollen datasets.}
 \vspace{-0.5cm}
 \label{fig:epoch}
\end{figure}

Furthermore, we demonstrate how the clustering performance metrics ARI and NMI change with the iterative contrastive learning process on the muraro and pollen datasets in Fig.~\ref{fig:epoch}. We can see that in the early epochs (epoch $<$60 on muraro and epoch $<$50 on pollen), both metrics undergo a rapid increasing and acutely fluctuation period. After that, the metrics enter a relatively stable period. Certainly, excessive training will also lead to overfitting and result in slight degradation of model performance, as we can see on the pollen dataset.

\subsection{Ablation Study}
Here, we conduct ablation studies on the effect of cluster-aware contrastive learning.

\begin{figure}[htbp]
  \centering
  \vspace{-0.2cm} 
  \includegraphics[width=0.75\textwidth]{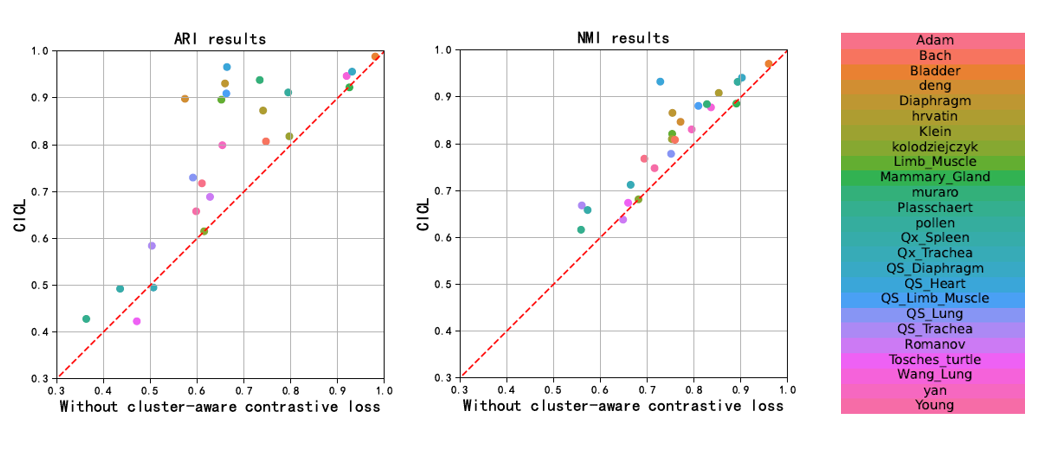}
  \vspace{-0.1cm}
  \caption[comparison]{Comparison between CICL and a variant without the cluster-aware contrastive loss. Vertical axis indicates results of our method, and horizontal axis stands for the results of the variant model. Each point in the figures represents the result on a dataset. Points above the $y=x$ line indicate that our method outperforms the variant on the corresponding datasets.}
  \label{fig:Ablation}
  \vspace{-0.4cm}
\end{figure}

\textbf{Cluster-aware contrastive loss.} 
One of the major innovations of CICL is the cluster-aware contrastive learning mechanism, which incorporates cluster structure information into contrastive loss, thereby enhancing the representations of cells. 
To validate the effectiveness of our mechanism, we conduct an ablation study. For comparison, we consider a variant without the cluster-aware loss (i.e., the 2nd term in Equ.~(\ref{equ:loss})). 
The results are presented in Fig.~\ref{fig:Ablation}. Here, the vertical axis shows the results of our method, and the horizontal axis presents the results of the variant without the cluster-aware contrastive loss. Notably, in terms of both ARI and NMI, the majority of points lie above the line $y = x$, indicating that CICL outperforms the variant model, which affirms the efficacy of our new contrastive learning loss. Nevertheless, we also see that on some datasets~(e.g. kolodziejczyk, Mammary\_Gland and Tosches\_turtle), CICL exhibits similar or even inferior performance. This is possibly caused by the errors of pseudo-labels.  

\begin{figure}[htbp]
  \centering
  \vspace{-0cm} 
  \includegraphics[width=0.4\textwidth]{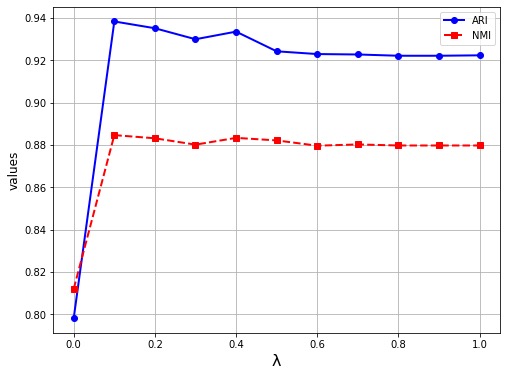}
  \caption[lambda]{The ARI and NMI results of CICL on muraro under different  $\lambda$ values.}
  \label{fig:lambda}
  \vspace{-0.5cm}
\end{figure}

\textbf{Effect of hyperparameter $\lambda$.} Here, we investigate the effect of the hyperparameter $\lambda$ on the performance of the model. We increase $\lambda$ from 0 to 1.0, and report the performance in terms of ARI and NMI. The results are illustrated in Fig.~\ref{fig:lambda}. We can see that our method has the worst performance at $\lambda = 0$ (without the cluster-aware contrastive loss). As $\lambda$ increases, the performance is improved rapidly. When $\lambda = 0.1$, both ARI and NMI reach the highest point. After that, ARI and NMI decrease slightly and gradually tend to be stable with the increase of $\lambda$. The result indicates that the model considerably benefits from the cluster-aware contrastive loss. 
In our experiments, we set $\lambda$=0.1, to avoid potential negative impact of pseudo-label error.

\section{Conclusion}
In this paper, to boost the performance of scRNA-seq data clustering analysis, we propose a novel approach called CICL. CICL adopts an iterative representation learning and clustering framework with an innovative cluster-aware contrastive loss. By comprehensively exploiting the underlying cluster structure of the training data, CICL can learn better scRNA-seq data presentations and thus achieve better clustering performance progressively. 
%
%
Extensive experiments on 25 real scRNA-seq datasets show that CICL outperforms the state-of-the-art methods in most cases, and achieves dominate advantage over the existing methods on average. Future work will focus on replacing K-means with advanced clustering methods to generate accurate pseudo-labels, and extending our idea to other downstream scRNA-seq data analysis tasks.

\bibliographystyle{plainnat}
%
\bibliography{ref}


\end{document}